\RequirePackage{fix-cm}
\documentclass[twocolumn,natbib]{svjour3}          
\smartqed  
\usepackage{graphicx}
\usepackage{float}
\usepackage{subfigure}
\usepackage{epstopdf, epsfig}
\usepackage{color}
\usepackage{sidecap}
\usepackage{amsfonts} 
\usepackage{amssymb}
\usepackage{hyperref}
 \journalname{Experiments in Fluids}
\begin{document}

\title{Measurement of interfacial wave dynamics in orbitally shaken cylindrical containers using ultrasound pulse-echo techniques  
}

\titlerunning{Interfacial wave dynamics in orbitally shaken containers}        

\author{Gerrit Maik Horstmann         \and
	    Markus Wylega                 \and
        Tom Weier 
}


\institute{Gerrit Maik Horstmann \at
              Helmholtz-Zentrum Dresden - Rossendorf \\
              Bautzner Landstr. 400, 01328 Dresden, Germany \\
              \email{g.horstmann@hzdr.de}           
           \and
           Markus Wylega \at
           Helmholtz-Zentrum Dresden - Rossendorf \\
           Bautzner Landstr. 400, 01328 Dresden, Germany \\
           \email{m.wylega@hzdr.de}
           \and
           Tom Weier \at
              Helmholtz-Zentrum Dresden - Rossendorf \\
              Bautzner Landstr. 400, 01328 Dresden, Germany \\
              \email{t.weier@hzdr.de}
}

\date{Received: date / Accepted: date}

\maketitle

\begin{abstract}
We present a novel experiment on interfacial wave dynamics in orbitally shaken cylindrical vessels containing two- and three fluid layers. The experiment was designed as a hydrodynamical model for both aluminum reduction cells and liquid metal batteries to gain new insights into the  rotational wave motion driven by the metal pad roll instability. Different options are presented to realize stable and measurable multi-layer stratifications. We introduce a new acoustic measurement procedure allowing to reconstruct wave amplitudes also in opaque liquids by tracking ultrasonic pulse echoes reflected on the interfaces. Measurements of resonance curves and phase shifts were conducted for varying interface positions. A strong influence of the top and bottom walls were observed, considerably reducing wave amplitudes and eigenfrequencies, when the interface is getting close. Finally, measured resonance curves were successfully compared with an existing forced wave theory that we extended to two-layer interfacial waves. The comparison stresses the importance to carefully control the boundary condition at the contact line.
\end{abstract}

\section{Introduction}
In the recent years, the investigation of rotational free-surface and interfacial waves has increasingly come to the fore of scientific awareness in connection with different research fields. Great attention to rotational interfacial waves is paid particularly in the field of Magnetohydrodynamics (MHD). 
Different kinds of instability mechanism has been identified being able to excite potentially dangerous interfacial waves in two important devices: firstly, in Hall-H\'{e}roult aluminum reduction cells (ARC)s, which are widely employed for aluminum production and secondly, in liquid metal batteries (LMB)s, which are discussed today as a cheap grid scale energy storage. An ARC consists of a stable density stratification of a molten salt mixture (cryolite) floating atop liquid aluminum. A strong cell current is applied to a graphite anode during the production process in order to reduce Al$_2$O$_3$ (dissolved in the cryolite) to Al. An extensive overview of the Hall-H\'{e}roult process can be found in \cite{Evans2007}. The cryolite-aluminum interface is well known to be highly susceptible to Lorentz-force driven instabilities. Most prominent is the so-called metal pad roll (MPR) instability firstly described by \cite{Sele1977}. He identified the Lorentz force arising due to the interaction of horizontal cell currents with external magnetic fields as the driving mechanism causing rotational interfacial waves. These waves limit the minimal height of the cryolite layer that has a crucial impact on the energy consumption. For that reason, the MPR was intensively studied analytically and numerically in the last decades \citep{Sneyd1992,Sneyd1994,Bojarevics1994,Davidson1998,Lukyanov2001} and still is subject of current research \citep{Pedchenko2009,Molokov2011,Pedcenko2017,Hua2018}.\\
In the last few years, the MPR was further found to be a crucial limiting factor for the safe operation of physically similar LMBs. A LMB consists of three stably stratified, inmiscible liquids: a low density alkaline or earth-alkaline liquid metal is floating on the top, a heavy metal is placed on the bottom and a molten salt mixture (electrolyte) of medium density is sandwiched in between. The electrolyte is ion-conductive for cations of the light metals, which are electrochemically reduced in the lower layer during discharge. An detailed review about the concepts and applicability of LMBs is given by \cite{Kim2013b}.
Very high cell currents can appear during cell operation especially in large-size LMBs, making the batteries vulnerable to different kinds of fluid dynamical instabilities. The resulting liquid flows can at worst disrupt the electrolyte leading to a short-circuit, or at best increase the mass transport in the battery. Extensive overviews of various instabilities in LMBs and their consequences can be found in \cite{Weier2017, Kelley2018}. Just as for ARCs, the MPR instability was identified as a key instability mechanism capable to trigger high-amplitude interfacial waves in large LMBs, limiting the maximum cell size. In contrast to ARCs, two interfaces are present in LMBs making stability predictions far more complex. For some LMBs, as the magnesium-antimony cell \citep{Weber2017c,Weber2017,Bojarevics2017}, the MPR behaves exactly in the same manner as in ARCs, since there the interfaces can be considered as decoupled and wave motion is present only at one interface. However, for other combinations of active materials, both interfaces can be simultaneously displaced, either symmetrically in phase or antisymmetrically phase-shifted by $180^{\circ}$, as recently investigated by \cite{Horstmann2018,Zikanov2018,Tucs2018,Molokov2018}. Both modes differently affects the stability of LMBs and are currently subject to intensive research.\\
The most serious issue of all the efforts to understand Lorentz force-driven interfacial instabilities is the lack of any experimental validation. Almost all studies have explored the MPR instability numerically or analytically since the realization of lab-size MHD experiments is very difficult. Cell currents, sufficiently strong to trigger interfacial waves, can be realized only by maintaining high operation temperatures for practicable active materials in both two- and three-layer systems, preventing the deployment of commercial measurement techniques as Ultrasound Doppler Velocimetry (UDV). Further, many active materials are chemically aggressive causing unwanted side reaction difficult to control. Only \cite{Pedchenko2009} presented a simplified ARC model experiment, where the electrolyte layer was replaced by a grid of thin vertical rods mimicking the characteristic vertical current distribution in liquid electrolytes. While the electromagnetic driving mechanism of the MPR instability could be successfully reproduced, it is quite likely that the many rods have a considerable influence on the hydrodynamics. This gave us the motivation to construct a complementary model experiment, which circumvents the MHD driving mechanism, but allows us to study the hydrodynamical response of the system, the resulting interfacial wave motion and coupling dynamics.    \\
Our idea was to replace the rotating Lorentz force distribution, which drives the MPR, by a rotating centripetal force, which can excite exactly the same rotational wave mode commonly associated with the MPR. The easiest way to apply a horizontal rotating centripetal force is to fix some fluid-filled observation sample onto an orbital shaker, which performs homogeneous horizontal translations along fixed circular paths of constant angular velocity.\\
The responding rotational sloshing dynamics in such systems were intensively studied in the last years, since there is a class of associated one-layer experiments in the context of another research field: most recently, the fluid dynamics in orbitally shaken bioreactors has been extensively studied with increasingly strong interest.  
Bioreactors are widely used for various applications as the cultivation of mammalian stem cells, drug production or fermentation processes since they can provide mixing, aeration, and shear stresses on multiple scales.
An overview about the opportunities and limitations of orbitally shaken bioreactors is given by \cite{Kloeckner2014}. These bioreactors are of particular interest for fluid dynamists since they offer a high potential for optimization in terms of mixing efficiency and power consumption. Various properties as the influence of different characterized wave and flow regimes to the mixing efficiency, the mass transport or shear stress distributions were numerically and experimentally studied by \cite{Weheliye2012,Ducci2014,Reclari2014, Bouvard2017, Weheliye2018, Alpresa2018a,Alpresa2018b}. Some of their findings can be directly transferred to interfacial wave systems, as we will discuss in Sect. \ref{sec:Theory}, and are interesting for the MPR wave as well. \\
As an advancement of the free-surface experiments listed above, we present the first multi-layer interfacial orbital shaking experiment in the paper at hand, which allows precise simultaneous measurements of interfacial displacements and phase shifts in both two- and three layer systems. The experiment was employed to pursue multiple objectives. Operating with two layers, it represents a hydromechanical model for ARCs. Here, it is of particular interest to study viscous damping rates in dependence of the viscosities, contact line dynamics and the interface position. Dissipation rates of standing interfacial waves are still widely unexplored but very important to predict stability onsets of the MPR instability. Using three stratified layers, we gain a model-experiment to study interfacial wave dynamics in LMBs. To predict the stability of LMBs, it is of major importance to understand the coupling of both interfaces, as the amplitude ratios and phase shifts between both waves. As shown by \cite{Horstmann2018}, the interfacial wave coupling in LMBs is in very good approximation determined only by pure hydrodynamical pressure exchanges, Lorentz forces are negligible for the coupling. For that reason, the experiment is a promising model to better understand the complex fluid mechanics in LMBs. Further, we are also interested in fundamental research of forced waves dynamics, since the three-layer system is a very descriptive example for a forced, non-linear and coupled oscillator. Finally, the experiment was also employed as a forerunner for a MHD experiment we pursue as a long-time goal. In operation with electrically conductive fluids, mechanically excited waves will probably decay slower when a electrical current is applied, which allows to extract the growth rates associated with the MPR instability. For this purpose, we employed an acoustic measurement technique, allowing to measure interfacial elevations also in opaque liquid metals, instead of optical techniques commonly applied in shaking experiments.\\
In Sect. \ref{sec:exp} we describe the experimental set-up and show various possibilities to create stable and measurable two- and three-layer stratifications by combining different silicon oils with paraffin oils and water. \\
In Sect. \ref{sec:Theory} we describe and expand existing forced linear wave theory of orbitally shaken waves to two-layer systems and justify that the linear approach is even more relevant for slower oscillating interfacial waves in comparison with free-surface waves. \\
A novel measurement procedure is presented and validated in Sect. \ref{sec:Procedure} that allows us to precisely measure interfacial elevations and phase shifts by directly tracking ultrasound echoes reflected at the interfaces. \\
Finally, we present and discuss first experimental results in Sect. \ref{sec:Results}. Resonance curves and phase shifts were measured for different interfacial positions in the first part. In the second part, we compare our measurements to the theory and discuss the influence of different contact line boundary conditions.  

\section{Experimental implementation}
\label{sec:exp}
\subsection{Experimental set-up}
\label{sec:set-up}
\begin{figure*}
	\setlength{\unitlength}{1cm}
	\begin{picture}(10,14)
	\put(1,13.5){\large \textbf{(a)}}
	\put(0,0){
		\def\svgwidth{270pt}    
		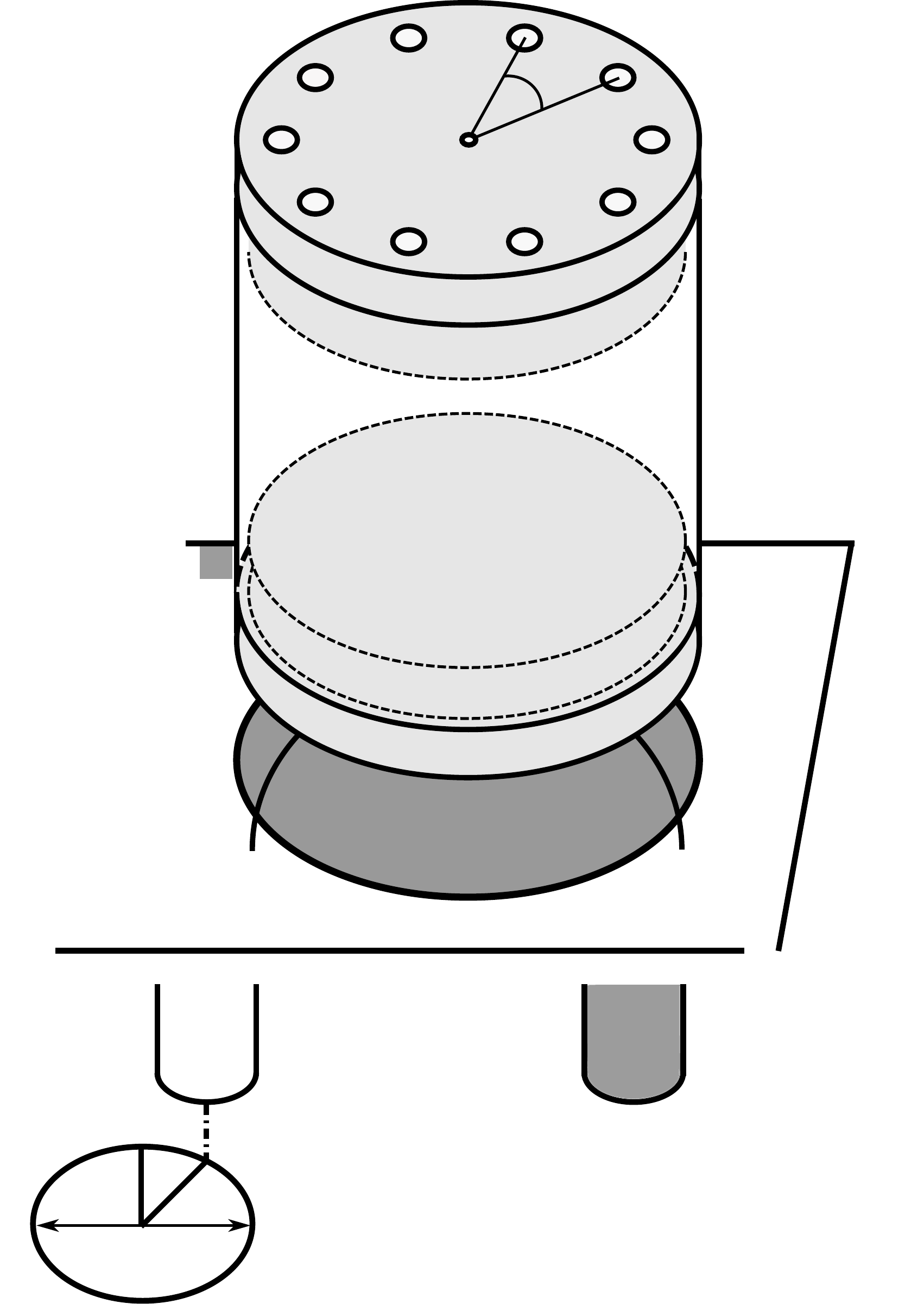
	}
	\put(11.2,13.5){\large \textbf{(b)}}
	\put(11.2,7.2){\includegraphics[width=0.29\textwidth]{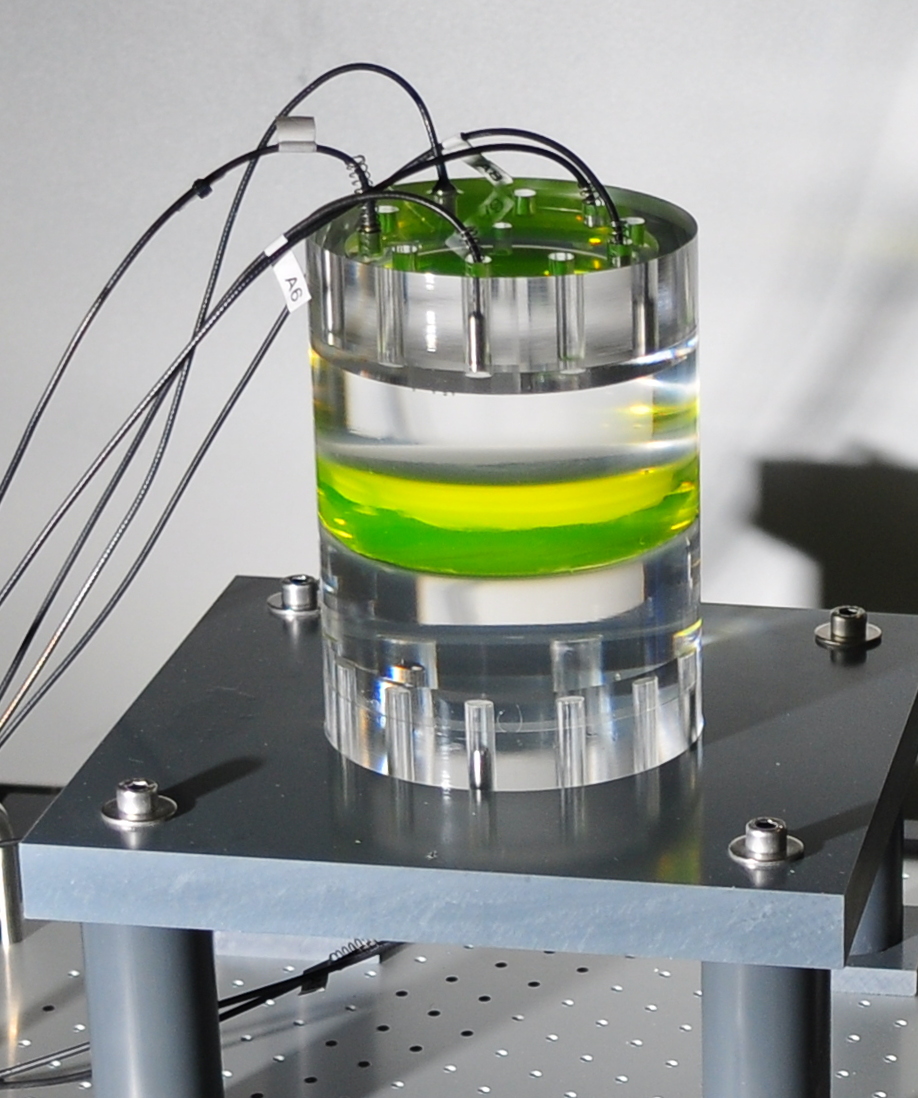}}
	\put(11.2,6.55){\large \textbf{(c)}}
	\put(9.85,0.15){
		\def\svgwidth{177pt}    
		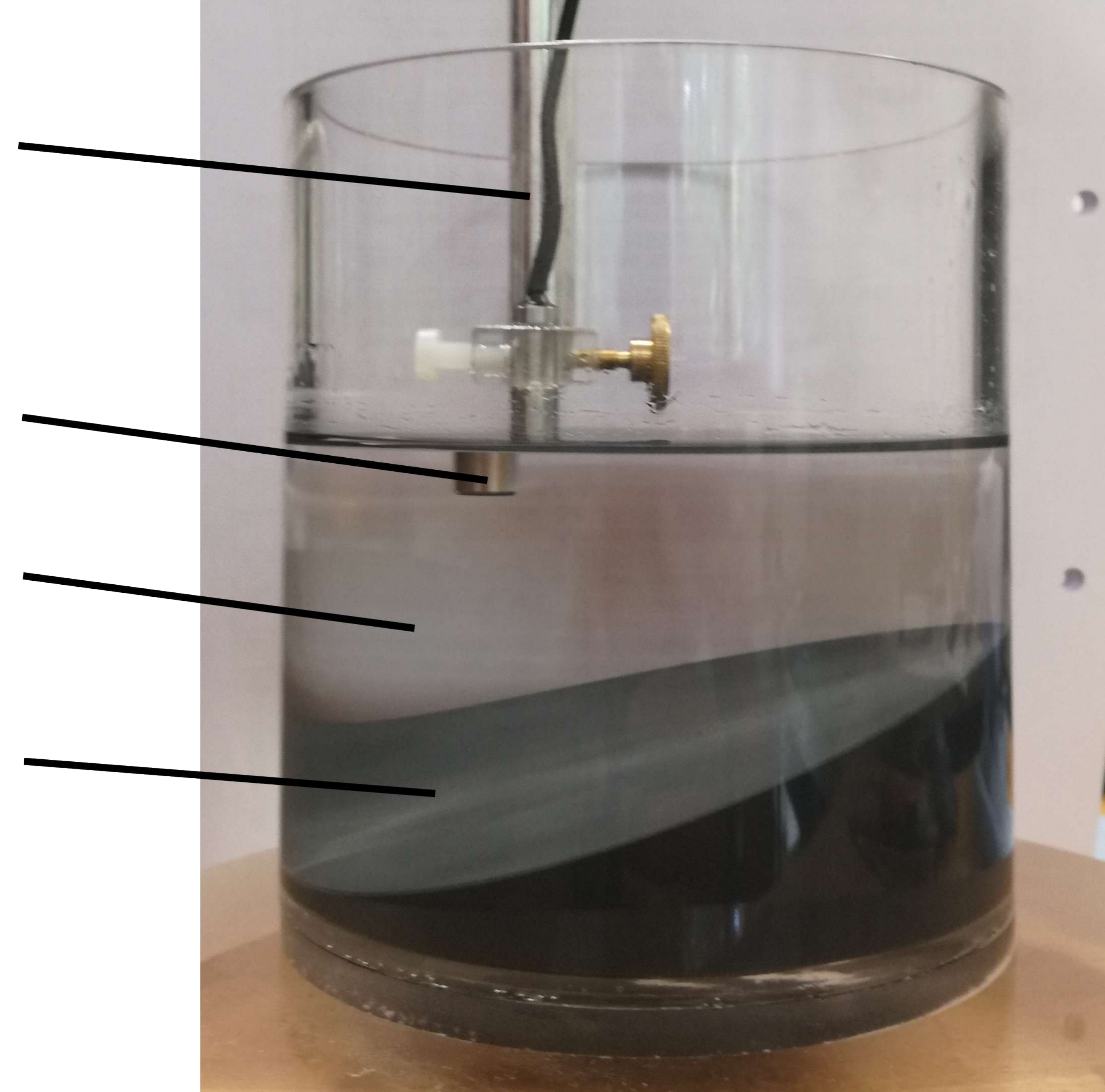
	}
	\end{picture}
	\caption{\textbf{(a)} Schematic illustration of the cylindrical observation volume placed on the shaking table. $(a)$ filling hole, $(b)$ UDV probe connection sockets, $(c)$ top lid, $(d)$ hollow cylinder, $(e)$ bottom lid, $(f)$ peg, $(g)$ stand, $(h)$ UDV probe plugs. The entire system is mounted on a modified Kuhner LS-X shaking table (not shown here) performing circular oscillations of fixed radius $d_s$ at a constant frequency $\Omega$. \textbf{(b)} photograph of the cylindrical cell filled with three inmiscible liquids: high-dense silicon oil Wacker\textsuperscript{\textregistered} AR 200 at the bottom, low-dense silicon oil Wacker\textsuperscript{\textregistered} AK 5 at the top and pure water colored with fluorescein in between. Different UDV probes are plugged in the cylindrical lids. \textbf{(c)} photograph of the reference device for phase-shift measurements, (a) crossbar, (b) UDV probe, (c) working fluid and (d) PVC cylinder with tilted top-face.}
	\label{fig:setup}
\end{figure*}
The experimental set-up was designed with the aim to facilitate simultaneous measurements of coupled interface motion even in opaque fluids.
A schematic illustration of the set-up is shown in Fig. \ref{fig:setup}\textbf{(a)}. Measurements were conducted in cylindrical samples realized by enclosing hollow cylinders (d) with differently sized top and bottom lids (c) and (e). Two hollow cylinders with different internal diameters $D = 100\, {\rm mm}, 200\, {\rm mm}$ and the same height of $H_c=120\, {\rm mm}$ were drilled from poly(methyl methacrylate) (PMMA) and polished. Aiming to make the experiment flexible for specific needs as well as to realize different aspect ratios, various lids were constructed with different internal heights $H_1 , H_2 = 10\, {\rm mm}, 35\, {\rm mm}, 60\, {\rm mm}$. The internal height $H$ of the observation volume is then given by $H = H_c - H_1 - H_2$ and can be adapted to the values $H = 100\, {\rm mm}, 75\, {\rm mm}, 50\, {\rm mm}$ yielding the aspect ratios $\Gamma = H/D = 1, 0.75, 0.5$ for the small and $\Gamma = 0.5, 0.375, 0.25$ for the large cylinder. The flat lids $H_1 , H_2 = 10\, {\rm mm}$ additionally contain ten equally distributed sockets (b) of $8\, {\rm mm}$ diameter to attach UDV probes. The distance between the central point of the lid and the central point of the sockets is $R_h = 42\, {\rm mm}$ for the small and $R_h = 90\, {\rm mm}$ for the large cylinder. To ensure completely non-invasive measurements, the UDV probes are not in direct contact with the working fluids. A $5\, {\rm mm}$ thick base was kept between the probes and the observation volume. All the lids were also made from PMMA chosen due to its transparency and beneficial acoustic properties for measurements of room temperature liquid metals as GaInSn. The top lids contain in addition a small filling hole (a) with a diameter of $4\, {\rm mm}$ enabling to degas the cell and to adjust the height of the fluid interfaces without the need to open the vessel. All components of the observation frame were manufactured with highest accuracy maintaining fabrication tolerances smaller than $0.05\, {\rm mm}$ in the internal dimensions. The entire cylindrical cell is placed on a stand that allows UDV measurements also from the bottom. For that, ten plugs (h), ideally suited to match the UDV sockets of the bottom lid (h), are introduced in the stand. Further, a peg (f) is installed to prevent own-turning motion of the cylinder. This whole configuration is mounted on a orbital Kuhner LS-X lab-shaker ($420\times420\, {\rm mm}$) not sketched here. This shaking table is able to perform ideal circular osculations $\varphi(t) = \Omega t$ of constant angular frequency $\Omega$ with a fixed shaking diameter $d_s$. The shaker was modified to allow a continuous adjustment of the shaking diameter $d_s$ up to $d_s = 70\, {\rm mm}$ and shaking frequencies $f = \Omega/2\pi$ from $f = 20\, {\rm rpm}$ up to $f = 500\, {\rm rpm}$.\\
For the acoustic measurement of the interfacial wave motions we used the ultrasound Doppler velocimeter DOP 3010 from Signal-Processing. The velocimeter can drive up to ten ultrasound probes, which produce beeps and listen for echoes. Usually, the measurement technique UDV is used to determine axial one-dimensional velocities out of the Doppler shift between emitted and received signals. However, we provide in Sect. \ref{sec:Procedure} a new procedure, which can determine wave amplitudes with a high accuracy by directly tracking the ultrasound echoes. So, strictly spoken, we are not conducting UDV in this paper. Up to twenty ultrasound probes, operating with $4\, {\rm MHz}$ and encompassing piezoelectric transducers of $5\, {\rm mm}$ diameter, are available for the measurements. By employing two DOP P3010 velocimeters, up to ten probes each at the top and at the bottom lid can be used simultaneously. This allows us to reconstruct interfacial wave modes even in three-layer systems at the both present interfaces simultaneously. All probes were lubricated with ultrasonic gel before inserted into the cylinder to optimize the acoustic coupling with the vessel. 
\begin{SCtable*}
	\caption{Different investigated working oils with the properties density $\rho$, kinematic viscosity $\nu$, surface tension $\gamma$ and sound velocity $c$ at room temperature. All paraffin oils can be combined with all silicon oils and all oils are combinable with water.}
	\label{tab:oils}       
	\begin{tabular}{lllll}
		\hline\noalign{\smallskip}
		working oil & $\rho$ (g/cm$^3$) & $\nu$ (mm$^2$/s)& $\gamma$ (mN/m)& $c$ (m/s) \\
		\noalign{\smallskip}\hline\noalign{\smallskip}
		Paraffinum Perliquidum (PP) & 0.846 & 36 & $\approx$30 & 1420.6 \\
		Paraffinum Subliquidum (PS)& 0.860 & 224 & $\approx$30 & 1455.5 \\
		Wacker\textsuperscript{\textregistered} silicone oil AK 5 & 0.92 & 5 & 19.2 & 980.4 \\
		Wacker\textsuperscript{\textregistered} silicone oil AK 35 & 0.955 & 35 & 20.7 & 998.4 \\
		Wacker\textsuperscript{\textregistered} silicone oil AR 200 & 1.04 & 200 & - & 1129.5 \\
		\noalign{\smallskip}\hline
	\end{tabular}
\end{SCtable*}
\subsection{Realization of multi-layer interfacial wave systems}
\label{sec:Oils}
Aiming to study interfacial wave dynamics, the most challenging part is to create suitable stably stratified multi-layer fluid systems within the observation cylinders. For that, appropriate working liquids have to be found. Possible working liquids must meet many different conditions restricting the number of liquids that can be applied. The following conditions must be fulfilled for this experiment:
\begin{itemize}
	\item All involved liquids must be inmiscible and nonreactive to realize stable interfaces
	\item The eigenfrequencies of the interfacial waves must be in the excitable range of the shaker $20 - 500\, {\rm rpm}$. This condition mainly restricts the fluid densities of the different phases that should not be too close to each other, see Sect. \ref{sec:Results}.
	\item The acoustic impedances of the single fluids must differ sufficiently strong to ensure satisfactory ultrasound reflection at the interfaces. This restriction is caused by the chosen measurement technique, however, since it will be sufficient when either the sound velocity or the density differ from each other, this restriction limits possible working fluids only slightly.
	\item The working liquids need to be very pure such that all physical properties as the densities and the viscosities remain homogeneous to ensure reproducibility.
	\item The viscosities must be chosen on the one hand small enough to excite visible wave motion but on the other hand high enough to avoid resonance catastrophes. The range of possible viscosities is directly determined by the range of possible shaking diameters $d_s$. For that reason the employed Kuhner shaker LS-X was specially modified to allow high shaking diameters up to $d_s = 70\, {\rm mm}$.
	\item Depending on the measurement objectives and cell sizes, also interfacial tension must be taken into account. Interfacial tension can influence the natural frequencies, cause a meniscus and lead to contact line pinning effects at the side wall that can in turn increase the damping rates. 
\end{itemize}
Many different combinations of liquids were tested regarding the above mentioned conditions. We found that different combinations of paraffin oils, silicon oils and water are suitable to create stable two- and three-layer systems. Table \ref{tab:oils} shows different liquid oils we found to fulfill all the above conditions (apart from interfacial tension effects that are difficult to control). The given physical properties were measured at room temperature or taken from technical data sheets. We found, that all given oils can be combined with water and all paraffin oils can be combined with all silicon oils to realize two-layer systems. Only the combinations PP$|$AK 5 and water$|$AR 200 transpired to be partially soluble, however, first indications of mixing were observed after a few days such that even long measurement series are nevertheless practicable. We found for all combinations different contact line dynamics that need to be further studied depending on an individual's motivation. However, we observed the general tendency that paraffin oil - water combinations lead to a considerable meniscus and almost fixed contact lines at the PMMA wall. These systems involve similar boundary conditions as surface waves in brimful cylinders and can be used to verify the theories by \cite{Miles1991} and \cite{Henderson1994} transferred to interfacial waves. On the contrary, for most paraffin oil - silicone oil combinations we found almost free-moving contact lines and a contact angle of almost $\Theta \approx 90^{\circ}$. This condition is usually assumed in common gravity-capillary wave theories (see Sect. \ref{sec:Theory}) such that these systems are very suitable for comparisons. \\
It is further possible to form many different three-layer systems involving two interfaces using the listed oils and water. For example, the lower-dense silicone oils AK 5 and AK 35 can be sandwiched between one of the paraffin oils floating at the top and water sitting at the bottom. When using the high-dense silicon oil AR 200, water can also form the middle layer by filling paraffin oil or a AK silicone oil on the top. All in all, we have found eight working combinations providing plenty of scope to modify the system parameters. As an example, Fig. \ref{fig:setup} \textbf{(b)} shows a photograph of the small cylindrical cell filled with the phase combination AR 200$|$water$|$Ak 5. The water layer is colored with fluorescein to increase the contrast between the three phases. Thereby, two sharp interfaces become clearly visible.\\ Finally, we also have found partially stable four-layer systems by using the silicon oil AR 200 as the bottom phase and sandwiching a AK silicon oil between water and paraffin oil. Four-phase wave dynamics are probably a bit too exotic, but such systems are rare in literature and might be of general interest, especially since all provided liquids are low-cost and non-toxic. 
\subsection{Realization of phase-shift measurements using a reference device}
One interesting physical feature of excited interfacial motion are phase-shifts between the oscillator and the excited interfacial waves, which can arise in dependence of the excitation frequency $\Omega$. Measurements of phase-shifts can be principally facilitated e.g. by tracking the orbital motions of the shaking table using light barriers and synchronizing the signal with the ultrasound signals. However, due to some benefits clarified later, we facilitated phase-shift measurements by employing a simple reference device, shown in Fig. \ref{fig:setup} \textbf{(c)}, that is placed beside the shaking table in the frame of reference at rest. It is made of a acrylic glass hollow cylinder (internal diameter $D = 9.2\, {\rm cm}$) that is closed at the bottom and can be filled with an arbitrary liquid from Table \ref{tab:oils}. In addition, different truncated polyvinylchlorid (PVC) cylinders were fabricated that can be inserted into the cylinder. The top-faces are tilted planes of different inclinations defined by the gradient angle $\alpha$. Phase shift measurement can be realized by immersing a UDV probe, that is mechanically connected to the shaking table via a crossbar, into the working liquid. During shaking operation, the probe follows the circular motion of the shaking table resulting in a circular motion of the probe (marked by the blue arrow) in
the reference frame of the resting device. The distance between the inclined top-plane and the probe, that can be measured by tracking ultrasound echoes, is then subject to an exact harmonic oscillation analytically described by a simple cosine function. By comparing this oscillation with the interfacial wave motion at fixed points, we can easily reconstruct the phase-shift as explained in Sect. \ref{sec:Procedure}. Moreover, the reference device was used to validate our measurement procedures and evaluations, since the oscillation amplitude of the probe-plane distance is directly given by the shaking diameter $d_s$ in dependence of the gradient angle $\alpha$. Using that relation, the device can also be exploited to measure sound velocities. Two differently tilted PVC cylinders with $\alpha = 19.4^{\circ}$ and $29.6^{\circ}$ are available and were selected for each measurement in dependence on the adjusted shaking diameters $d_s$.
\section{Theoretical description of orbitally forced wave motion}
\label{sec:Theory}
The first theoretical models of forced wave dynamics in moving containers are traced back to the beginning of the space exploration age. In that time it was of great interest to understand the sloshing dynamics of spacecraft propellants and their resulting forces on the tank walls. Using Potential flow theory, wave and sloshing dynamics in free surface liquids were analytically described in laterally shaken tanks of various different geometries \citep{Abramson1966}. Despite the fact that the assumptions needed for Potential flow theory are drastic (inviscid, incompressible and irrotational), remarkable agreement with many experiments conducted later was found in an astonishing high range of the parameter space \citep{Ibrahim2005}. Surprisingly, the fluid dynamics of orbitally shaken containers were not investigated in those times. A Potential model of orbitally excited free-surface gravity waves in cylindrical containers was presented first very recently by \cite{Reclari2014}. They found, using cylindrical coordinates ($r, \varphi, z$), the following form for the free surface elevation $\eta (r, \varphi , t)$:
\begin{eqnarray}
\label{eq:Wave}
&&\eta(r,\varphi,t) = \frac{d_s \Omega^2}{2g}\cos(\Omega t - \varphi)\nonumber \\
&&\times\left\{r + \sum_{n=1}^{\infty}\left[\frac{D}{(\epsilon_{1n}^2 -1)}\frac{\Omega^2}{(\omega_{1n}^2 -\Omega^2)}\frac{J_1 ( \frac{2\epsilon_{1n}r}{D})}{J_{1}(\epsilon_{1n})}\right]\right\}.
\end{eqnarray}
Here, $d_s$ denotes the shaking diameter, $\Omega$ the angular shaking frequency, $g$ the standard acceleration due to gravity and $D$ the inner diameter of the cylindrical tank, whereas $J_1$ is the Bessels's function of first kind and first order, and $\epsilon_{1n}$ are the radial wave numbers, calculated as the $n$ roots of the first derivate of $J_1$. Finally, $\omega_{1n}$ are the wave number-dependent natural eigenfrequencies of gravity waves given by the dispersion relation
\begin{equation}
\omega_{1n}^{2} = \frac{2g\epsilon_{1n}}{D}\tanh\left(\frac{2\epsilon_{1n}h_1}{D} \right),
\label{eq:1LayerDisp}
\end{equation}
where $h_1$ denotes the filling height of the working liquid.
The wave amplitude $\eta_0$ can be calculated from Eq. \ref{eq:Wave} by setting $\eta_0 = \eta (R, \varphi = \Omega \tilde{t}, t=\tilde{t})$.
Please note that Eq. \ref{eq:Wave} was derived using linear boundary conditions such that it is valid only for sufficiently small wave amplitudes. The first modes of weakly nonlinear wave modes were also derived and are described in detail by \cite{Reclari2013}. \\
From Eq. \ref{eq:Wave} some interesting features of orbitally driven surface waves can be recognized. At first, it is interesting to note that the orbital
shaking motion excites only non-axisymmetric wave modes $n \in \mathbb{N}_1$ in the linear regime. In general, free gravity waves in cylindrical tanks are determined by two degrees of freedom $\epsilon_{mn}$, where $m \in \mathbb{N}_0$ (described by the Bessel's function of $m$-th order) is the number of nodal diameters and $n \in \mathbb{N}_1$ the number of nodal circles, respectively. 
The first wave mode ($m,n=1,1$) is the mode widely associated with the MPR and of most interest for the present study.
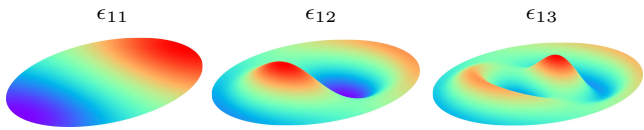
\begin{figure}
	\def\svgwidth{238pt}    
\begingroup%
  \makeatletter%
  \providecommand\color[2][]{%
    \errmessage{(Inkscape) Color is used for the text in Inkscape, but the package 'color.sty' is not loaded}%
    \renewcommand\color[2][]{}%
  }%
  \providecommand\transparent[1]{%
    \errmessage{(Inkscape) Transparency is used (non-zero) for the text in Inkscape, but the package 'transparent.sty' is not loaded}%
    \renewcommand\transparent[1]{}%
  }%
  \providecommand\rotatebox[2]{#2}%
  \newcommand*\fsize{\dimexpr\f@size pt\relax}%
  \newcommand*\lineheight[1]{\fontsize{\fsize}{#1\fsize}\selectfont}%
  \ifx\svgwidth\undefined%
    \setlength{\unitlength}{1395.18181801bp}%
    \ifx\svgscale\undefined%
      \relax%
    \else%
      \setlength{\unitlength}{\unitlength * \real{\svgscale}}%
    \fi%
  \else%
    \setlength{\unitlength}{\svgwidth}%
  \fi%
  \global\let\svgwidth\undefined%
  \global\let\svgscale\undefined%
  \makeatother%
  \begin{picture}(1,0.19582969)%
    \lineheight{1}%
    \setlength\tabcolsep{0pt}%
    \put(0,0){\includegraphics[width=\unitlength,page=1]{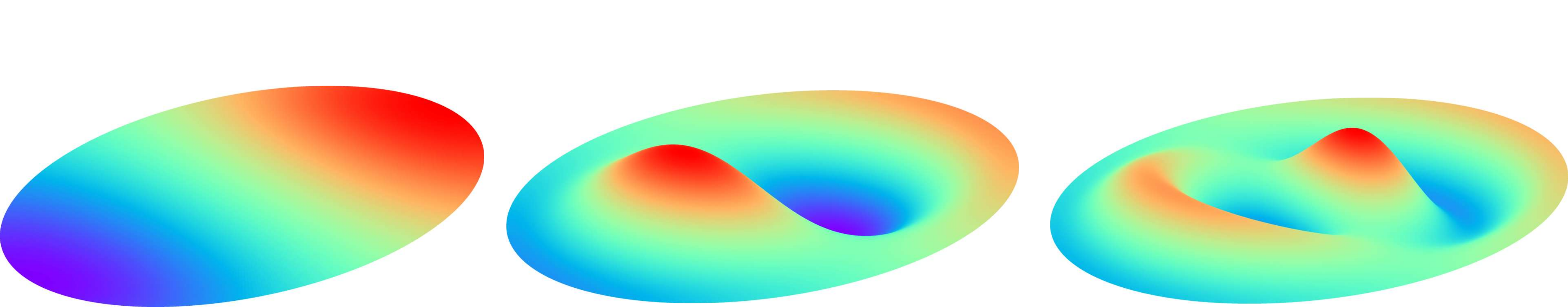}}%
    \put(0.11803117,0.20567703){\color[rgb]{0,0,0}\makebox(0,0)[lt]{\begin{minipage}{0.13669494\unitlength}\raggedright \end{minipage}}}%
    \put(0.14106964,0.17035138){\color[rgb]{0,0,0}\makebox(0,0)[lt]{\lineheight{1.20000005}\smash{\begin{tabular}[t]{l}$\epsilon_{11}$\end{tabular}}}}%
    \put(0.46898387,0.17035138){\color[rgb]{0,0,0}\makebox(0,0)[lt]{\lineheight{1.20000005}\smash{\begin{tabular}[t]{l}$\epsilon_{12}$\end{tabular}}}}%
    \put(0.81625044,0.17035138){\color[rgb]{0,0,0}\makebox(0,0)[lt]{\lineheight{1.20000005}\smash{\begin{tabular}[t]{l}$\epsilon_{13}$\end{tabular}}}}%
  \end{picture}%
\endgroup%
	
	\caption{Visualization of the first three non-axisymmetric linear wave modes $\epsilon_{11}$, $\epsilon_{12}$ and $\epsilon_{13}$ due to Eq. \ref{eq:Wave} of an exemplarily chosen, fixed amplitude. The mode $\epsilon_{11}$ is commonly associated with the MPR instability.}
	\label{fig:Modes}
\end{figure}
For a clearer understanding, Fig. \ref{fig:Modes} shows the first three modes in conformity with Eq. \ref{eq:Wave} of an exemplarily chosen, fixed amplitude. It can be seen how the number of crests and troughs is increased in the lateral direction with increasing $n$, while in the direction of rotation the number of crests remains constant on every cycle ($m=1$). However, that result is not anymore valid in the nonlinear regimes, where also $m\neq 1$ and sub-harmonic wave modes can be found \citep{Reclari2014}.\\
Secondly, in Eq. \ref{eq:Wave} the wave amplitude diverges when the shaking frequency approaches the eigenfrequencies $\omega_{1n}$ of the free surface. This phenomenon is well known as the resonance catastrophe where the wave absorbs a maximum amount of the energy provided by the exciter. However, an infinite wave amplitude is unphysical and due to the fact that resisting damping forces were completely neglected in the inviscid Potential model. Consequently, the theory can only predict wave motion sufficiently far away from the eigenfrequencies where viscous damping effects are of minor importances. \\
Thirdly, always when $\Omega$ exceeds the eigenfrequencies, a sign change in the wave amplitude can be obtained. That can be attributed to an occurring phase shift between the shaking table and the waves at the resonance point. For $\Omega < \omega_{1n}$ wave and shaker oscillate perfectly in phase, while for $\Omega > \omega_{1n}$ Eq. \ref{eq:Wave} predicts a phase delay $\Delta$  of of $\Delta = 180\, ^{\circ}$ between the forcing and the wave motion. In that regime the liquid distribution at the tank wall is in opposition to the inertial forces. That phenomenon is also well known from orbitally shaken bioreactors, where the phase shift was found to significantly influence the flow fields and the mixing of the flow \citep{Weheliye2018}. This predicted phase shift could be verified by different experiments in the regions sufficiently far away from the resonance frequency. Near resonance, viscous damping has to be considered which smooths the phase jump around the resonance, as clearly visible in \cite{Bouvard2017}, where silicon oils of relatively high viscosities are used.\\   
The presented model accounts only for the free surface in one-phase fluids. Though, the main objective of our experiment is to investigate interfacial wave dynamics in two- and three-layer liquids. To our best knowledge, orbitally shaken multi-layer systems have not yet been described in the literature. For that reason we transferred the Potential model of \cite{Reclari2014} to two-phase fluids following the same procedure. Surprisingly, we found that Eq. \ref{eq:Wave} and all its presented properties also exactly hold true in two-layer fluids, with the sole difference that the dispersion relation changes to
\begin{equation}
\omega_{1n}^{2} = \frac{2(\rho_2 - \rho_1)g\frac{\epsilon_{1n}}{D}}{\rho_1\coth(2\frac{\epsilon_{1n}}{D}h_1) + \rho_2\coth(2\frac{\epsilon_{1n}}{D}h_2)},
\label{eq:2LayerDisp}
\end{equation} 
where the index 1 denotes the upper liquid and the index 2 the lower liquid of higher density, respectively. The two-layer eigenfrequencies are noticeably lower in comparison with the one-layer frequencies (Eq. \ref{eq:1LayerDisp}). From this follows that the presence of a second liquid layer damps the wave motion in the sense that lower wave amplitudes are reached around the resonance in comparison with one-layer systems for the same chosen shaking diameters $d_s$. This is because the wave elevation scales with $\sim \Omega^2$ and resonance happens in two-layer fluids for smaller values of $\Omega$. This property is beneficial for the application of linear wave theory, since the parameter space ($d_s$,$\Omega$) of the linear regime is considerably increased. This argumentation is also underlined by the regime classification study conducted by \cite{Alpresa2018a}. They pointed out the importance of the Froude number (Fr), showing that linear waves are only excitable for sufficiently small Fr. In multi-layer systems, we always find lower Fr for the same chosen $d_s$ in comparison with one-layer systems, since the restoring gravity force is smaller due to smaller density jumps at the interface. We further emphasize the higher applicability of linear wave theory in Sect. \ref{sec:Results}, where the presented resonance curves comprise very small amplitude ratios even for comparatively large chosen $d_s$.\\
The free interfacial wave dynamics of three-layer liquids involving hydrodynamically coupled interfaces is theoretically studied by \cite{Horstmann2018} in the context of liquid metal batteries and found to be far more complex. The description of forced three-layer wave motion is even more complicated and out of the scope of the present study. 
\section{Experimental procedure and reconstruction of interfacial waves}
\label{sec:Procedure}
\begin{figure}
	\hspace*{-0.3cm}
	\def\svgwidth{244pt}    
	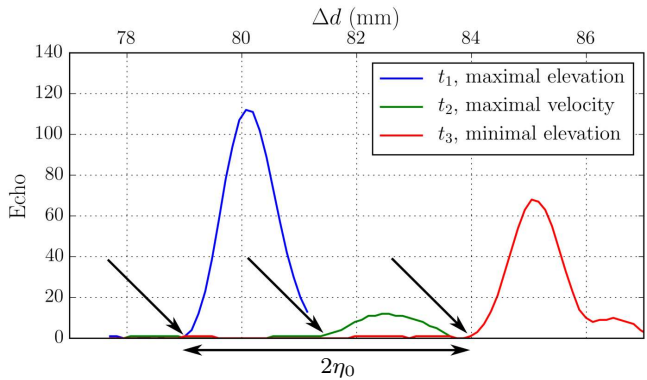	
	\caption{Measured echoes at the distance $\Delta d$ from the UDV probe reflected at the AK35$|$PP interface ($\Omega = 33\, {\rm rpm}$, $d_s = 2.5\, {\rm cm}$, $h_1 = 2\, {\rm cm}$) for three exemplarily chosen times: $t_1$ maximal interfacial elevation, $t_2$ position of maximal vertical velocity and $t_3$ minimal interfacial elevation. The three black arrows mark the used tracking points in the emergence area of the echoes. The distance between the tracking points of minimal and maximal elevation corresponds to the double amplitude of the interfacial wave $2\eta_0$.}
	\label{fig:Echoes}
\end{figure}
Since the experiment was designed to fasciculate measurements of interfacial movements also in opaque liquids, we had to establish an acoustic measurement technique. A widespread technique established for measurements in liquid metals is the Ultrasound Doppler Velocimetry (UDV), commonly used to measure one dimensional (1D) axial velocities in direction of implemented UDV probes. The probes emit consecutively short ultrasonic pulses, which are reflected on existing or artificially added scattering particles in the working liquid. After some transit time depending on the distance of the single scattering particles, the pulse echoes are recaptured the probes. By means of differences in transit time of the scattering echoes between consecutive pulse emissions, resulting from a finite shift of the particle position, the movement of the scattering particles and consequently the flow velocity in axial direction can be determined. In principal, interfacial elevations can be reconstructed by measuring interfacial velocities in vertical direction. However, we found that interfacial elevations can be measured with far higher accuracy by tracking the pulse echo of the interface directly.\\
\begin{figure}
	\vspace*{-0.37cm}
	\hspace*{-0.3cm}
	\def\svgwidth{244pt}    
	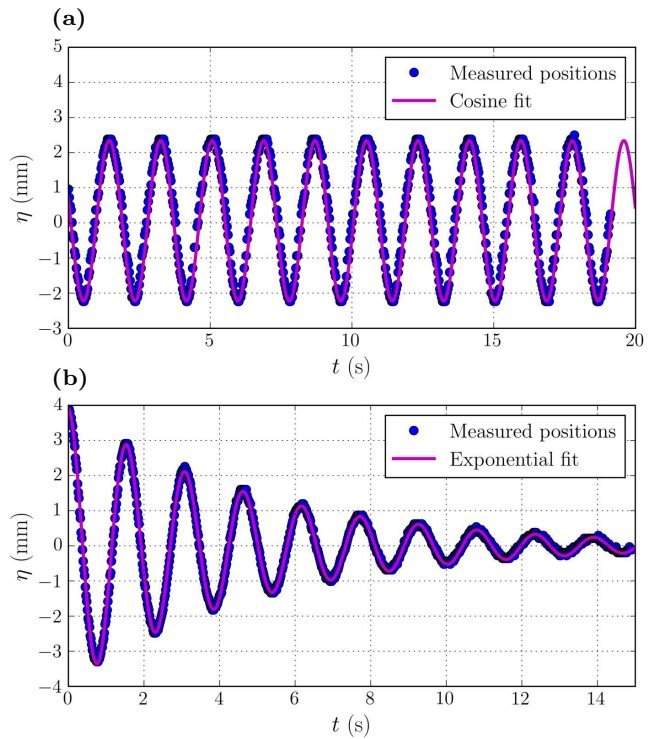
	\caption{\textbf{(a)} interface elevation $\eta$ reconstructed by tracking the constant echo value ${\rm echo}=4$ from Fig. \ref{fig:Echoes} plotted over time $t$. A cosine fit is shown as well. \textbf{(b)} reconstructed interface elevation $\eta$ of Ak 5$|$water ($\Omega = 39\, {\rm rpm}$, $d_s = 1.25\, {\rm cm}$, $h_1 = 5\, {\rm cm}$) over time after switching off the shaking table. An exponential fit yields a damping rate of $\gamma = 0.201 \pm 0.013\, 1/{\rm s}$.}
	\label{fig:Oscillation}
\end{figure}
To allow an adequate tracking, it has to be ensured that the echo of the interfaces is significantly stronger than echoes caused by impurities or gas bubbles. For that reason all the working liquids were degassed before use and the high care was taken during the filling processes. Under those conditions a very clear echo of the interface can be extracted by optimizing the ultrasound parameters as the burst length and sound intensity. Fig. \ref{fig:Echoes} shows the normalized, non-dimensional echo of the Ak 35$|$PP interface at three different, exemplarily chosen times: at maximal interface elevation ($t_1$), minimal interface elevation ($t_3$) and maximal vertical velocity at the rest position ($t_2$). During the measurements, the echo is moving from the left (blue curve) to the right (red curve) and back again with the exciting frequency $\Omega$, where the distance between the minimal and maximal position reflects the double wave amplitude $2\eta_0$. Hence, the interfacial movement can be principally reconstructed by tracking the echo positions. However, as visible in Fig. \ref{fig:Echoes}, the echo is not maintaining its shape throughout the movement. Especially, when the interface is in rapid motion, it becomes considerably flattened (green curve). Moreover, we observed splits of the echo as well as unpredictable deformations in some measurements. These changes are caused by the changes of the interfacial orientation deflecting the echoes during the oscillation. Thereby, the amount of the ultrasound pulse that is reflected back to the UDV probe can be altered. Due to these issues, it is complicated to track the echo by means of the maximum value or the average position. Different methods were tested and compared. It turned out that tracking the first point of the evolving echo which is significantly higher than the background noise at the positions before facilitates reliable measurements of interfacial oscillations. These points are marked by the three black arrows for the three echo positions. They reflect the edge of the interface, the first spacial point assigned to the interface, where the ultrasound pulse is scattered. We observed, that the first few echo values, depending on the used liquids and ultrasound parameters, are not affected by the deformations of the echo and follow perfectly the expected harmonic motion of small amplitude waves. This is emphasized in Fig. \ref{fig:Oscillation} \textbf{(a)} showing the wave elevation $\eta$ over time, calculated by tracking the value ${\rm echo}=4$ in Fig. \ref{fig:Echoes}. Amplitude $\eta_0$ and phase $\varphi$ are determined by fitting a cosine function using non-linear least squares. The corresponding fit is shown in Fig. \ref{fig:Oscillation} \textbf{(a)} as well and agrees perfectly to the measured values. Using this method, we could adequately reconstruct wave amplitudes in the range $0.15\, {\rm mm} \lesssim \eta_0 \lesssim 10\, {\rm mm}$. The lower limit is determined by the spacial resolution ($0.1\, {\rm mm}$) of the ultrasound sampling intervals, while the upper limit is engendered by a too strong deflection of the echo for large wave amplitudes. For each measurement series the number of fitted oscillation periods was adapted depending on the signal quality and wave amplitudes to guarantee high quality fits with coefficients of determination larger than $R^2 > 0.97$.\\
The determination of amplitudes using cosine fits is limited to linear wave motion, however, this method can be further exploited to identify the transition from linear to weakly nonlinear waves.
\begin{figure}
	\centering	
	\hspace*{-0.28cm}
	\includegraphics[width=0.5\textwidth]{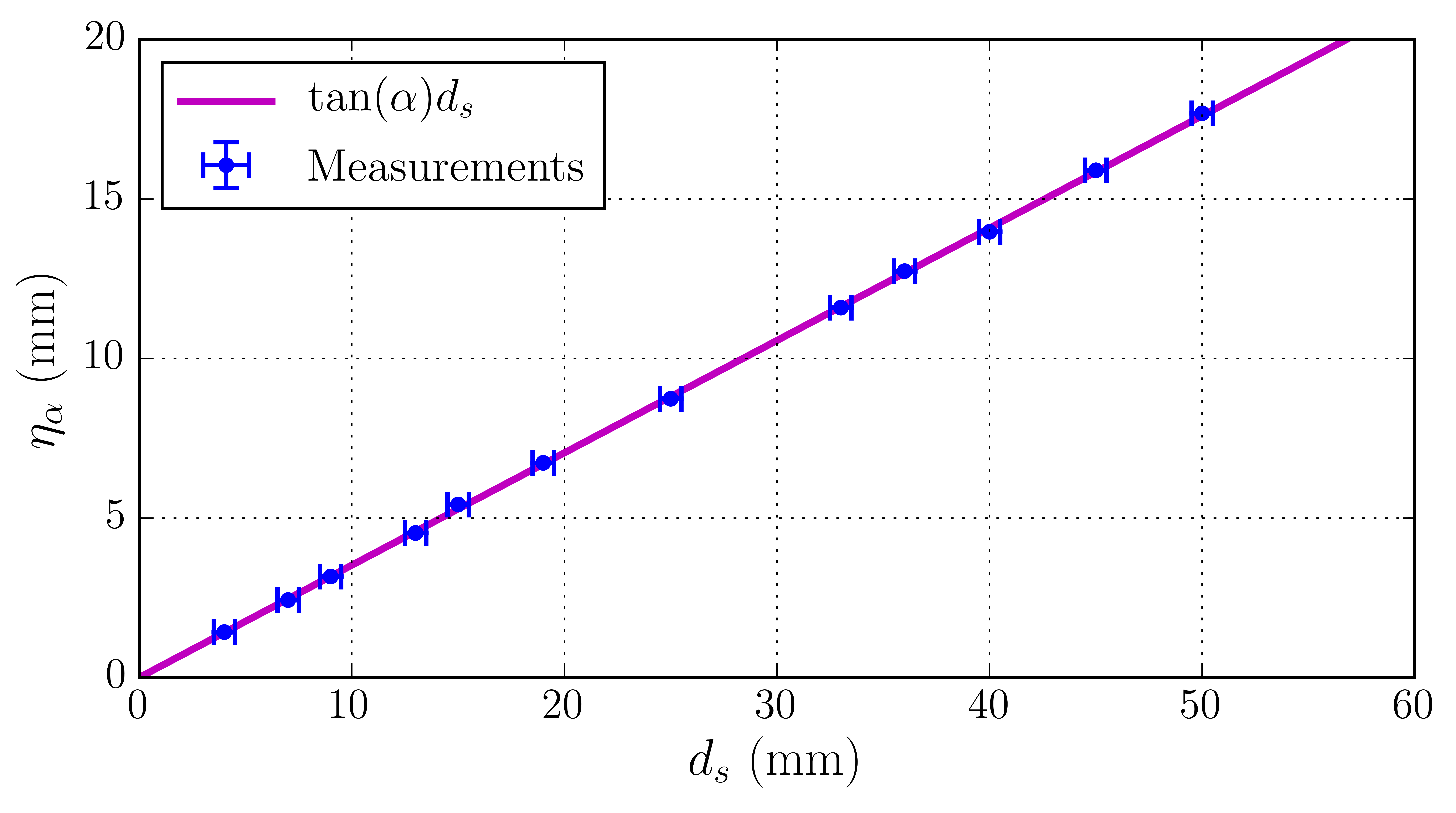}
	\caption{Measured oscillation amplitudes $\eta_{\alpha}$ in dependence of the shaking diameter $d_s$ in the reference device filled with PP.}
	\label{fig:Validation}
\end{figure}
Typically, the symmetry of the oscillatory motion will break, when the wave is approaching the nonlinear regime near to the resonance. The slope of the crests is often found to increase, finally causing wave breaks, while the troughs are flattened. Even slight deformations of that kind will immediately lead to drops of the coefficients of determination. For that reasons $R^2$ is an appropriate measure to determine the confines of the linear regime quantitatively and was used to ensure the applicability of linear wave theory to our measurements in Sect. \ref{sec:Results}. \\
Measurements of viscous damping rates $\gamma$ were realized in a similar way. Since the damping rates depends largely on the excited wave modes \citep{Case1956,Troy2005}, the interface must first be excited with the corresponding resonance frequencies. Subsequently, the shaker was switched off to realize free wave motion that decays exponentially due to viscous dissipation. At this point it is crucial to give the interface sufficient time to develop free wave motion, since the braking acceleration of the shaker initially disturbs the wave. After about three periods we observed perfect exponential decays of the wave amplitudes as exemplarily shown in Fig. \ref{fig:Oscillation} \textbf{(b)} for AK5$|$water. The damping rate can be determined by either fitting the function
\begin{equation}
\eta_{r_i, \varphi_i}(t) = \eta_0 e^{-\gamma t}\cos(\Omega t - \phi),
\end{equation}
where $r_i$ and $\varphi_i$ mark the position of the $i-th$ UDV probe, $\eta_0$ denotes the initial amplitude and $\phi$ denotes the phase angle, or by calculating logarithmic decrements. We found that the fit determines $\gamma$ more accurately, especially for more viscous cases, where the waves can decay within just a few periods. Though, despite the fact that we received negligibly small fitting parameters in almost every case, slight changes of the damping rates were observed when fitting only the first or last periods of the decaying process especially for the viscous cases. These deviations where taken into account for the error estimations. For the shown example we receive a damping rate of $\gamma = 0.201 \pm 0.013\, 1/{\rm s}$.\\
\begin{figure*}
	\centering
	\def\svgwidth{500pt}    
	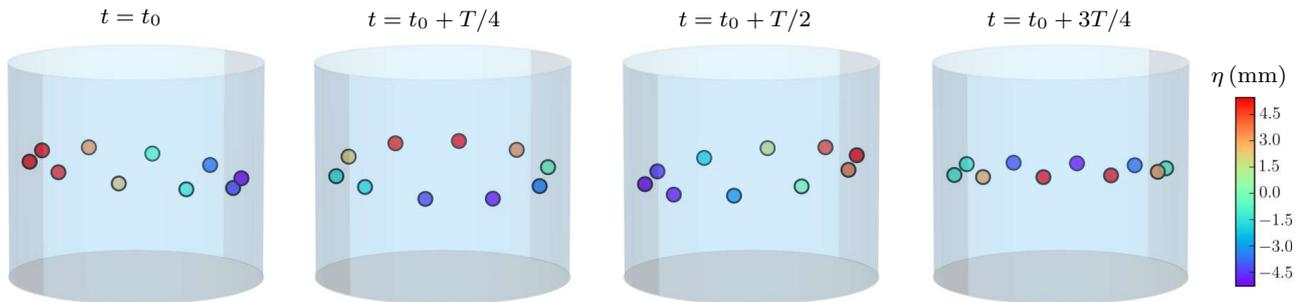	
	\caption{Reconstructed rotational wave mode $\epsilon_{11}$ of the PP$|$AK 35 interfaces located at the half height $\Gamma_{\eta} = 0.5\, {\rm cm}$ at different times within a period $T$. The wave was excited with $f = 39\, {\rm rpm}$. The reconstructed wave positions $\eta$ of the installed UDV probes are coded in color. A wave amplitude of $\eta_0 = 5,48 \pm 0.04\, {\rm mm}$ can be extracted by averaging over the probes.}
	\label{fig:RecWaves}
\end{figure*}
To improve the presented amplitude reconstruction procedure, further different filter techniques and error routines were implemented. A median average filter was used to smooth the echo signal. By that, the range of traceable echo values could be increased. In some measurements a low noise level before the interface echoes could not be maintained. Additional ghost peaks were sometimes found to occur that can be caused by small impurities in the fluids and on the interface or by multiple pulse reflections at the glass wall. To avoid trackings of false echo signals, an improved tracking routine was written that reduces the searching interval for the subsequent location of the interface position by using information out of its history. Since the interface is subject to finite accelerations, the searching interval $|\eta_{j+1} - \eta_{j}|$ for the next interface position $\eta_{j+1}$ was restricted by considering the distances of the previous $K$ consecutive tracking positions $\eta_{j-1}, \eta_{j-2},\cdots,\eta_{j-K}$ such that   
\begin{equation}
|\eta_{j+1} - \eta_{j}| < \chi \max \{|\eta_{j-k+1} - \eta_{j-k}|:k = 1,2,\cdots, K \}
\end{equation}
is fulfilled. The scaling factor $\chi$ is an empirical value chosen individually for each measurement in dependence of the fluid accelerations $\sim \Omega^2$. This simple method allowed us to reconstruct interfacial motion even in contaminated cases, e.g., where the interface was ruptured at resonance, as long as the first $K$ positions are tracked correctly. \\
Finally, the whole tracking procedure was validated since many sources of errors and heuristic parameters as the tracking echo are present. The reference devices are suitable for validation since the diameter $d \equiv d_s$ of the circulatory motion of the immersed UDV probe (Fig. \ref{fig:setup} \textbf{(c)}) is analytically connected via the gradient angle to the distance between probe and the tilted PVC cylinder. When the probe homogeneously rotates above the tilted flat bottom, its distance is subject to a perfect `harmonic' oscillation described by a cosine function, just as for linear interfacial wave motion. On that basis, the phase shift measurement are realized simply by comparing the fitted phases $\phi$ of the interfacial wave and the reference frame oscillation, since the UDV probes are synchronized. The oscillation amplitude $\eta_{\alpha}$ is given by the gradient angle $\alpha$: 
\begin{equation}
\eta_{\alpha} = \tan(\alpha)d_s .
\end{equation}
Using the described tracking procedure, oscillating amplitudes were measured  for different adjusted shaking diameters $d_s$. Fig. \ref{fig:Validation} shows the measured and theoretical oscillation amplitudes $\eta_{\alpha}$ for the working fluid PP and the small PVC cylinder with $\alpha=19.4\, ^{\circ}$. The shaking diameters cannot be specified directly and were measured after the adjustments with an uncertainty of $\sigma_{d_{s}} \approx 0.5\, {\rm mm}$. All measured values agree perfectly to the expected amplitude curve within the error intervals. We found, that the procedure is robust to changes of the used tracking echo or $\chi$, since the same oscillations can be reproduced in an extended range of chosen echo and $\chi$ values. Hence, our method is proven to track interfacial movements accurately.
\section{Results and discussion}
\label{sec:Results}
In the following we present and discuss interfacial wave dynamics in two-layer cases, since we found a very complex system behavior in the three-layer cases exceeding the scope of this study. In the first part we present the results of a large PP$|$AK 35 measurement series to study the resonance dynamics, phase shift dynamics and the damping behavior in dependence of the interface position. In the second part we discuss the influence of different contact line dynamics and compare measured wave amplitudes with the Potential model.
\subsection{Wave dynamics in dependence of the interface position}
\begin{figure}
	\centering	
	\hspace*{-0.28cm}
	\includegraphics[width=0.5\textwidth]{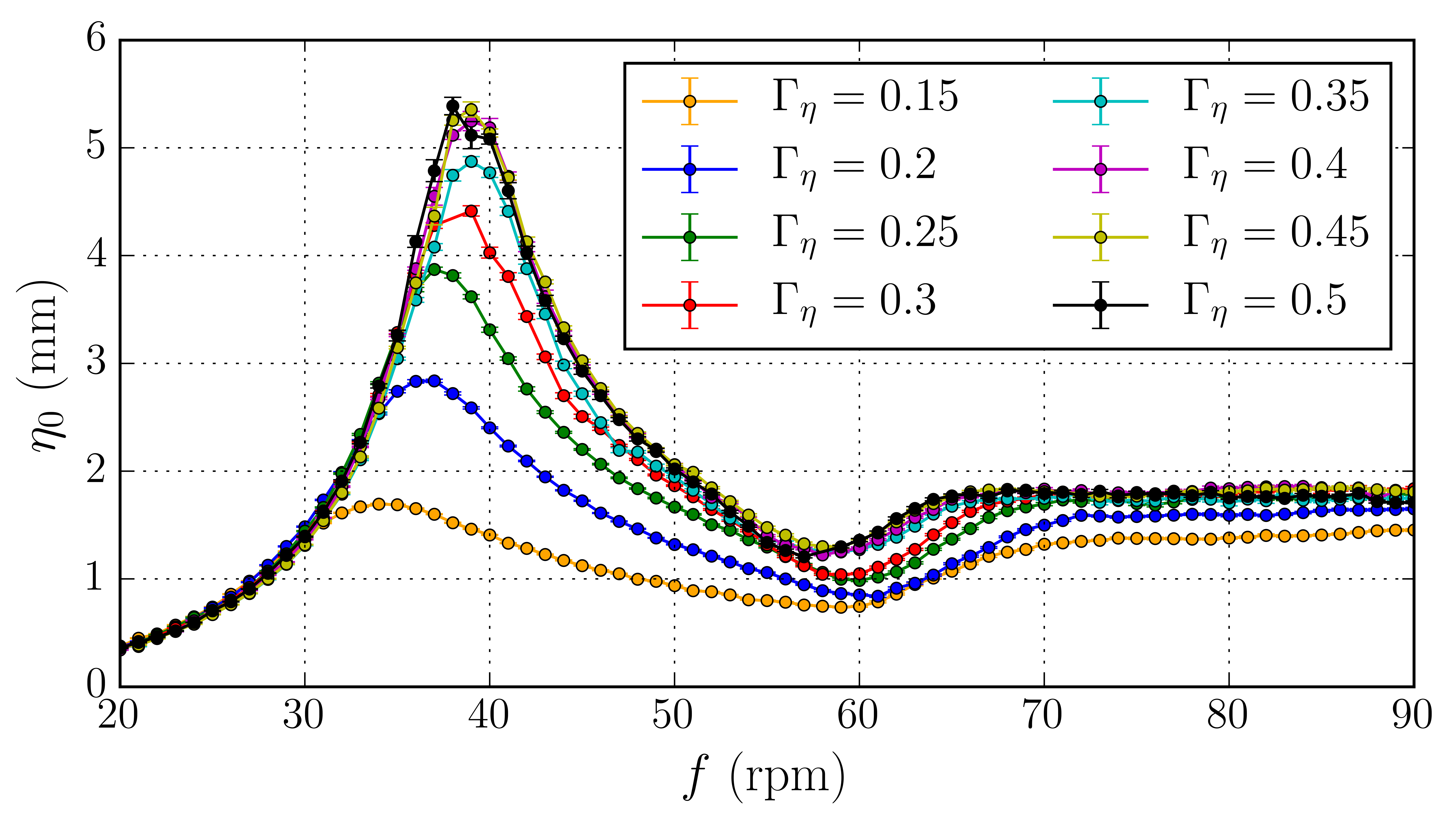}
	\caption{Resonance curves of PP$|$Ak 35 for different interface positions $\Gamma_{\eta}$ located in the lower half of the cylinder.}
	\label{fig:Low}
\end{figure}

It is known from one-layer surface waves that the height of the liquid layer can highly influence wave properties as the eigenfrequencies and damping rates, especially when the liquid layer becomes shallow. In two-layer systems the interface position has an even greater impact since there are two different wall regions (at the bottom and at the top) which can individually determine the overall wave dynamics in a different way. To get insights into the influence of the interface location, we measured resonance curves around the first mode $\epsilon_{11}$ in PP$|$AK 35 by specifying the shaking frequency from $20$ to $90\, {\rm rpm}$ for 15 different interfacial positions. The small cylinder $D = 10\, {\rm cm}$ with $\Gamma = 1$ was used and a shaking diameter of $d_s = 25\, {\rm mm}$ was adjusted for all measurements. We define the dimensionless interface position by $\Gamma_{\eta} = h_2 /H$ ranging from $\Gamma_{\eta} = 0$ (bottom) to $\Gamma_{\eta} = 1$ (top). $\Gamma_{\eta}$ was modified from $\Gamma_{\eta} = 0.15$ to $\Gamma_{\eta} = 0.85$ in steps of $\Gamma_{\eta} = 0.05$.\\
At first, Fig. \ref{fig:RecWaves} shows exemplarily the reconstructed three-dimensional interface elevation $\eta$ of the first mode $\epsilon_{11}$ by using ten UDV probes for the case $\Gamma_{\eta} = 0.5$. The current interfacial height at the ten sensor locations is shown for four chosen times in one wave period. Both, the shape of the wave $\epsilon_{11}$ as well as the clockwise rotary motion are clearly visible demonstrating that we can adequately reconstruct the complete spatio-temporal information of the first wave mode. To spatially reconstruct the shapes of higher modes involving more modal circles further radially distributed probe connection sockets are needed and will be manufactured for future studies. \\ 
Resonance curves were calculated by fitting the oscillation amplitudes $\eta_0$ for each single shaking frequencies. Fig. \ref{fig:Low} shows eight resonance curves corresponding to the measured interface positions located in the lower half ($0.15 \leq \Gamma_{\eta} \leq 0.5$) of the cylinder. This plot provides many insights into different physical mechanisms affecting the interfacial wave:
\begin{itemize}
	\item For each curve we can identify a peak of the $\eta_0$ marking the eigenfrequency of the first wave mode $\epsilon_{11}$.\\
	\item The eigenfrequency is decreased with decreasing $\Gamma_{\eta}$. While this effect is very small near the cell center, it becomes all the stronger when the interface is getting close to the bottom wall.\\
	\item The maximum peak amplitude decreases with decreasing $\Gamma_{\eta}$, again even stronger when the interface approaches the bottom wall.\\
	\item All resonance curves coincide for small shaking frequencies $f \lesssim 30\, {\rm rpm}$.\\
	\item All resonance curves reach a saturated regime for high shaking frequencies $f \gtrsim 70\, {\rm rpm}$. The saturation amplitude is decreased with decreasing $\Gamma_{\eta}$. 
\end{itemize}
Some of these behaviors can be readily explained by fundamental wave dynamics, while others have not been reported before. The presence of a distinct amplitude peak at the first eigenfrequency is obvious, since the wave can here naturally absorb the largest percentage of the energy injected by the orbital shaker. Due to the disregard of any dissipative friction forces, the Potential model Eq. \ref{eq:Wave} predicts an infinite amplitude at the resonance frequency. For the viscous liquids as used in this study the amplitudes near resonance are highly effected by friction forces. A finite resonance amplitude results out of the balance of the injected energy and dissipative energy losses.\\
\begin{figure}
	\centering
	\hspace*{-0.28cm}	
	\includegraphics[width=0.5\textwidth]{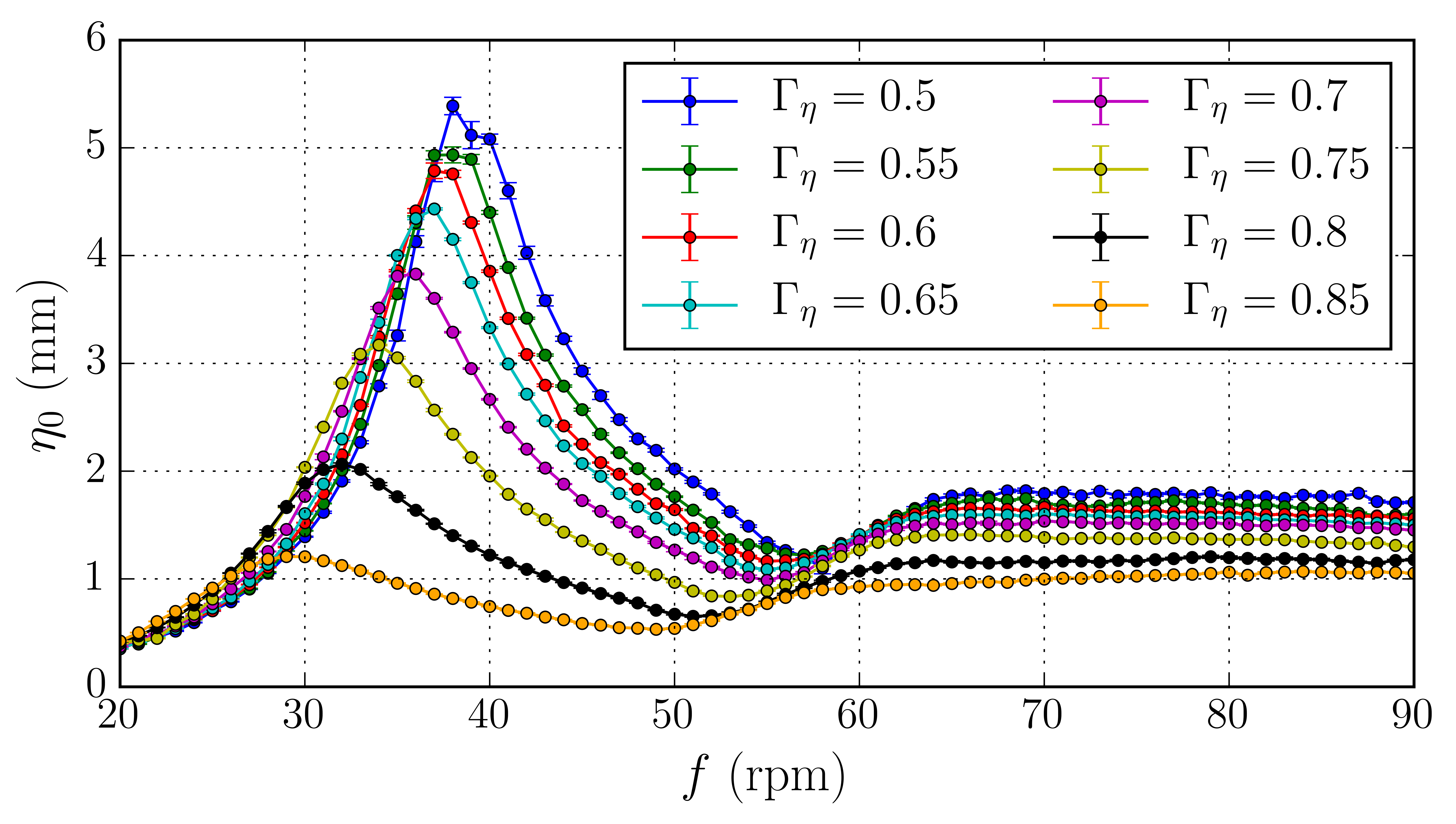}
	\caption{Resonance curves of PP$|$Ak 35 for different interface positions $\Gamma_{\eta}$ located in the upper half of the cylinder.}
	\label{fig:Up}
\end{figure}
The decrease of both the eigenfrequency and the resonance amplitude for decreasing $\Gamma_{\eta}$ can be explained by the interaction of at least two different mechanisms. On the one hand, the eigenfrequency naturally decreases when the interface approaches the bottom or top wall region as reflected in Eq. \ref{eq:2LayerDisp}. On the other hand, viscous dissipation is drastically increased when the interface is located near the bottom and top wall because then very high shear stresses will arise between the flow driving interface and the close wall, where the liquid is at rest. That phenomenon was very well described by \cite{Case1956} and \cite{Troy2005} who studied the contribution of different wall regions to the overall viscous damping rate. Moreover it is known that high damping rates also decrease the eigenfrequency of gravity waves further explaining the observed lowering of the eigenfrequency. Similarly, the reduction of the resonance amplitude can be clarified. Obviously, the increase of the damping rate reduces the maximum achievable amplitudes. But also the reduction of the eigenfrequency directly affects the amplitudes. It becomes readily apparent from Eq. \ref{eq:Wave} that the wave amplitude scales with $\Omega^2$. If now the eigenfrequency is reduced, the resonance point occurs for lower values of $\Omega$ also reducing the maximum wave amplitude. While all these phenomenons were principally known before, they have not been explicitly  reported in association with orbital shaking, since resonance curves were never studied in dependence of the interface position or the amplitudes near resonance already exceeded the linear regime and could not be resolved anymore \citep{Reclari2014,Bouvard2017,Alpresa2018a}. \\
The observation that the interfacial position and the resulting eigenfrequency does not influence the resonance curves for low shaking frequencies is already known from one-layer shaking experiments. In this regime the interfacial elevation is induced by a simple balance of the gravitational and centripetal pressure. A tilted non-wavy plane results out of the balance (described by the linear portion of Eq. \ref{eq:Wave}), where the slope was found to scale linearly with the Froude number \citep{Weheliye2012,Ducci2014}. \\
\begin{figure}
	\centering	
	\hspace*{-0.28cm}
	\includegraphics[width=0.5\textwidth]{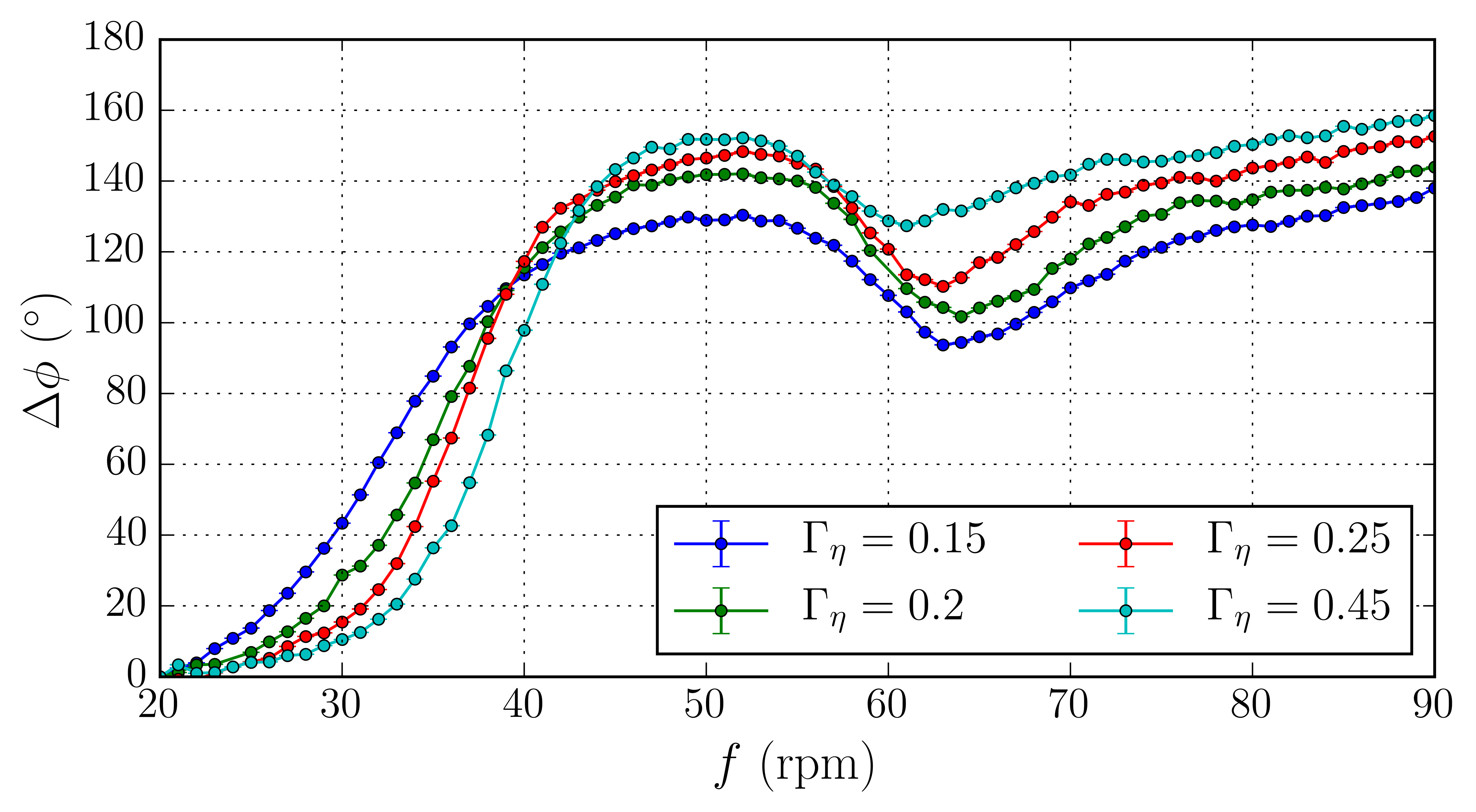}
	\caption{Phase shift $\Delta \phi$ between interface and shaker for different interface positions $\Gamma_{\eta}$ located in the lower half of the cylinder.}
	\label{fig:Phase}
\end{figure}
Finally, our explored saturation beyond the first wave mode is surprising at the first glance and was - to our best knowledge - never reported before. Usually, we had expected a second peak corresponding to the second wave mode $\epsilon_{12}$. However, we found that the system becomes overdamped in that region such that the peak is completely flattened out, since the viscous damping rate increases considerably for higher wave modes \citep{Troy2005}. We indeed investigated that the wave decays within one period for $f > 68$ after switching off the shaking table. \\
Complementary, Fig. \ref{fig:Up} shows the eight resonance curves of the interfaces located in the upper half ($0.5 \leq \Gamma_{\eta} \leq 0.85$) of the cylinder. In contrast to free surface waves, we expected a symmetric picture of decaying resonance curves, since the used oils have almost the same viscosity. While the resonance curves indeed behave qualitatively the same, the eigenfrequencies and resonance amplitudes decrease notably faster. These deviations may be caused by the slightly higher viscosity of PP or a small possible hysteresis of the contact line. Also the filling hole might slightly disturb the boundary layer at the top wall.
All in all, we can experimentally confirm a very large impact of both the top and bottom wall regions to the resonance dynamics. \\
Finally, we present the related PP$|$AK 35 phase shift measurements  $\Delta \phi$ between interface and shaker. Fig. \ref{fig:Phase} shows $\Delta \phi$ of four selected interface positions located in the lower half of the cylinder. In contrast to the sharp phase jump of $\Delta \phi = 180^{\circ}$ predicted by the non-viscous Potential model at the resonance frequency, a more complex and continuous phase shift behavior can be observed. This difference is produced by the relatively high viscosities of the used oils provoking high damping rates, which substantially smooth out the phase jump. This effect is well known from the damped harmonic oscillator under external excitation. For all shown interface positions the wave is (almost) in phase with the shaker for the smallest shaking frequencies $f$. The phase shift continuously increases until it reaches $\Delta \phi \approx 90^{\circ}$ at the resonance frequency and approaches to a maximum beyond at $f \approx 52\, {\rm rpm}$. As expected, the phase shift is all the more smoothed out for low $\Gamma_{\eta}$, since there higher damping rates are present. However, the most interesting fact we can observe here is that a phase shift of $\Delta \phi = 180^{\circ}$ is never reached, in contrast to the damped harmonic oscillator where the phase shift can be described by an arctangent function. A similar observation was recently made by \cite{Bouvard2017} for a high-viscous silicon oil. We assume, that the phase is smoothed out into the regime of the second wave mode $\epsilon_{12}$ where a second phase jump is predicted by Eq. \ref{eq:Wave}, such that the phase shift is reduced before the anti-phase flow can fully evolve. Indeed we find a decrease of the phase shift for $f > 52\, {\rm rpm}$ until the drop is suppressed when reaching the over damped regime. There, $\Delta \phi$ increases linearly for all curves with the same slope but different offsets. Such a phase shift behavior have not yet been reported and demands further research to understand the wave dynamics in the overdamped regime as it can be present especially in small orbitally shaken test tubes and flasks.
\subsection{Applicability of the Potential model}
\begin{figure}
	\hspace*{-0.2cm}
	\def\svgwidth{244pt}    
	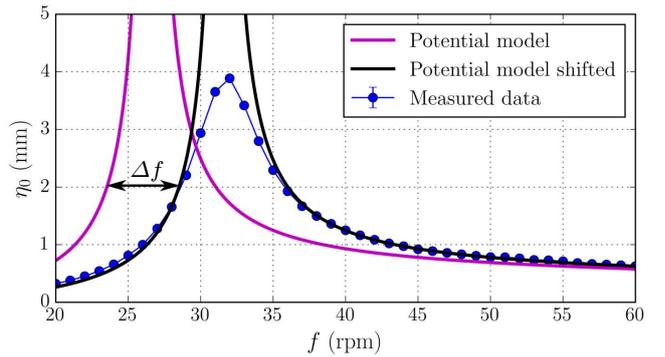	
	\caption{Resonance curve of the AK 35$|$water interface at the probe position $r=R_h = 42\, {\rm mm}$ with $D = 10\, {\rm cm}$, $\Gamma = 1$, $\Gamma_{\eta} = 0.5$ and $d_s = 34\, {\rm mm}$. The measured eigenfrequency is shifted from the theoretical gravity wave eigenfrequency by $\Delta f = 4.97\, {\rm rpm}$.}
	\label{fig:TheoAK35}
\end{figure}

\begin{figure}
	\hspace*{-0.2cm}
	\def\svgwidth{244pt}    
	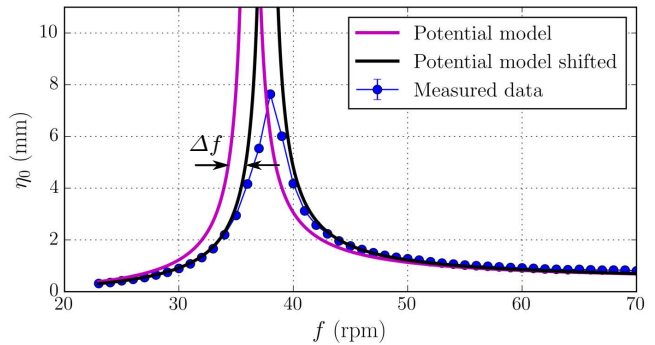	
	\caption{Resonance curve of the AK 5$|$water interface  at the probe position $r=R_h = 42\, {\rm mm}$ with $D = 10\, {\rm cm}$, $\Gamma = 1$, $\Gamma_{\eta} = 0.5$ and $d_s = 19\, {\rm mm}$. The side walls were coated with petroleum jelly. The measured eigenfrequency is shifted from the theoretical gravity wave eigenfrequency by $\Delta f = 1.7\, {\rm rpm}$.}
	\label{fig:TheoAK5}
\end{figure}

Finally, the measured resonance curves are compared with the adapted Potential model (Eq. \ref{eq:Wave}). The earlier presented PP$|$AK 35 measurement series is inappropriate for comparisons, since we observed relatively high damping rates but viscosity was completely neglected in the Potential model. For that reason we decided to use the AK 5$|$water and the AK 35$|$water systems for comparisons involving lower cumulative viscosities. Exemplarily, Fig. \ref{fig:TheoAK35} shows the AK 35$|$water resonance curve for $\Gamma_{\eta} = 0.5$. The measured amplitudes are shown as blue dots and the theoretical prediction is visualized by the magenta-colored line. The Potential model disagrees clearly with the measured amplitudes notably underestimating the resonance frequency. However, we found that the model describes the measured resonance curve almost perfectly, when the theoretical curve is simply shifted to the measured resonance frequency. This is highlighted by the black curve in Fig. \ref{fig:TheoAK35} showing the Potential prediction frequency-shifted by a factor of $\Delta f = 4.97\, {\rm rpm}$. The shifted curve agrees everywhere except for a small region near resonance, where the inviscid Potential theory diverges. The same behavior was found for all conducted silicon oil$|$ water measurements. The theory always underestimates the resonance frequency by a factor of $\Delta f \approx 2.5\, {\rm rpm}$ up to $\Delta f \approx 5\, {\rm rpm}$ but agrees remarkably after shifting the frequency. Hence, we concluded that the Potential model generally works well with the exception that the natural frequency of standing gravity waves (Eq. \ref{eq:2LayerDisp}) fails to account for our experimental conditions. The reason for this shift lies in the experimental boundary conditions of the contact line mentioned before in Sec. \ref{sec:Oils}. The Potential model accounts only for free slipping contact lines with a static contact angle of $90^{\circ}$ and ignores the influence of interfacial tension. This demands are in good approximation fulfilled in one-layer tanks filled with water, while the situation is rather complex for two-layer interfacial waves. The main reason for this is due to the fact that the wettability of different liquids at the tank wall can highly differ, which consequently leads to the evolution of distinct menisci or different contact line hysteresis effects. Indeed, we observed mostly fixed contact lines for all water$|$oil combinations. However, for larger amplitudes $\eta_0 \gtrsim 3\, {\rm mm}$ we could further recognize the formation of a thin film of oil penetrating the water layer resulting in a permanent displacement of the contact line. In this configuration the interface behaves as it slides almost unperturbed along the oil film rather than replacing oil on the wall with water, until it arrives the contact line where further elevation is abruptly suppressed. Exactly the same complex boundary dynamics were found in similar interfacial wave experiments but under vertical excitation before, see \cite{Ito1999, Ito2008, Batson2013}. \\
Despite the complexity of this contact line behavior we can quantitatively substantiate the frequency shift by two different effects. The first effect is simple. The relatively small density difference of the used liquids also result in a comparatively weak gravity force such that gravity not alone determines the wave dynamics. Also the capillary force contributes significantly to the overall restoring force such that we have to consider gravity-capillary waves rather than pure gravity waves. This can be easily justified by calculating the difference between the gravity-capillary wave eigenfrequency (given in \cite{Horstmann2018}) and the pure gravity wave eigenfrequency $\Delta f_{\gamma_{\eta}}$, given as
\begin{equation}
\label{eq:tension}
\Delta f_{\gamma_{\eta}} = \frac{1}{2\pi}\sqrt{\frac{8 \gamma_{\eta} \left(\frac{\epsilon_{1n}}{D}\right)^3}{\rho_1\coth(2\frac{\epsilon_{1n}}{D}h_1) + \rho_2\coth(2\frac{\epsilon_{1n}}{D}h_2)}},
\end{equation}
where $\gamma_{\eta}$ denotes the interfacial tension. For AK 35$|$water we measured a relatively strong interfacial tension of $\gamma_{\eta} = (81 \pm 0.6)\, {\rm mN}/{\rm m}$ resulting in a frequency shift due to capillarity of $\Delta f_{\gamma_{\eta}} = 3.25\, {\rm rpm}$.\\
The second effect is due to the partially fixed contact line, which effectively reduces the freely movable interfacial area increasing the natural frequency as well. The increase is quite complex to quantify for the partially fixed boundary condition caused by the oil film. However, for the purpose of a conservative estimate we can consider the frequency shift $\Delta f_{\rm pinning}$ caused by perfectly pinning contact line, which was derived by \cite{Miles1991} as well as \cite{Henderson1994} and is given for the first mode $\epsilon_{11}$ as  
\begin{equation}
\label{eq:pin}
\Delta f_{\rm pinning} =  \frac{\omega_{11}}{2\pi}\sqrt{\frac{8\left(\frac{\epsilon_{11}}{D}\right)^2 \gamma_{\eta}}{(\rho_2 -\rho_1)g(\epsilon_{11}^2 - 1)}},
\end{equation}
where $\omega_{11}$ is the angular frequency for free moving contact lines given by Eq. \ref{eq:2LayerDisp}. Please note, that the authors derived the pinning frequency for one-layer free-surface liquids. However, following their analysis, their expression can be easily expanded to two-layer interfacial waves by simply replacing the free-surface capillary length by the interfacial capillary length, which is reflected in Eq. \ref{eq:pin}. For AK 35$|$water we find a frequency shift due to contact line pinning of $\Delta f_{\rm pinning} = 2.7\, {\rm rpm}$, which can be understand as the highest possible shift, since the oil film boundary condition is expected to cause a frequency shift somewhere between the free moving and pinned contact line extrema. Hence, the total shift $\Delta f_{\rm film}$ can be found in the interval
\begin{equation}
\Delta f_{\gamma_{\eta}} \leq \Delta f_{\rm film} \leq \Delta f_{\gamma_{\eta}} + \Delta f_{\rm pinning}.
\end{equation}
That gives $3.25\, {\rm rpm} \leq \Delta f_{\rm film} \leq 5.95\, {\rm rpm}$, which indeed is in good agreement to the measured frequency shift of $\Delta f = 4.97\, {\rm rpm}$. Thus, the frequency shift, which appears high at the first glance, can be well explained by known interfacial tension forces acting on the interface and in the contact line region. Both shifts were not reported in comparable one-layer experiments, where the waves act as pure gravity waves in cylinders of this size and the contact line dynamics are easier to control. Therefore, the contact line boundary conditions has to be treated with utmost care in interfacial wave experiments and capillary forces have to be considered even in mid-size containers.\\
After identifying the crucial role of the contact line, it is worthwhile to manipulate the boundary condition aiming to better substantiate the general applicability of the Potential model in multi-layer interfacial wave systems.
For that reason we have tested different wall coatings trying to realize the free moving contact line boundary condition as good as possible. Different waxes as well as hydrophobic polytetrafluoroethylene (Teflon) were applied, which all influenced the contact angle significantly but nevertheless could not provide free contact lines. Finally, petroleum jelly, but only in combination with AK 5$|$water, brought temporarily the desired effect. In this set-up, no meniscus and only a very weak contact line hysteresis were observable. However, the petroleum jelly slowly dissolved in the silicon oil such that we could conduct reliable measurements only for a few hours. Fig. \ref{fig:TheoAK5} shows the AK 5$|$water resonance curve for $\Gamma_{\eta} = 0.5$ measured in the small cylinder $D = 10\, {\rm cm}$ coated with petroleum jelly. Now, the Potential prediction fits very well to the measured curve away from resonance, except for a small remaining frequency shift of $\Delta f = 1.7\, {\rm rpm}$. This remaining shift can be easily explained by the lack of interfacial tension. For AK 5$|$water we measured $\gamma_{\eta} = 35.1 \pm 0.5\, {\rm mN}/{\rm m}$ giving $\Delta f_{\gamma_{\eta}} = 1.1\, {\rm rpm}$ due to Eq. \ref{eq:tension}.
The small left shift may be related to experimental uncertainties or weak contact line hysteresis effects. Overall, these results finally confirm that the Potential model developed by \cite{Reclari2014} can be successfully applied also to interfacial waves, when the contact line boundary conditions are controlled. \\
As a last point, we want to emphasize once more the expansion of the linear interfacial wave regime in comparison with forced one-layer free-surface wave systems, naturally broadening the applicability of linear wave theory. \cite{Alpresa2018a} draw up a detailed regime classification of orbitally shaken free-surface waves. They provide a clear overview of all orbital shaking studies reviewed in the literature in cylindrical containers together with the applied system parameters. Nonlinear wave phenomena as breaking are reported near resonance in almost all publications, despite the fact most of them applied comparable or even lower shaking diameters in the range of $0.5\, {\rm mm} \lesssim d_s \lesssim 5\, {\rm mm}$. In contrast, we observed a completely linear and harmonic interfacial wave behavior even at resonance in almost all conducted measurements. One reason is surely the higher viscosity of the employed oils by contrast with water. But more important is the considerably lower eigenfrequency of multi-layer interfacial wave systems as discussed in Sec. \ref{sec:Theory}. Only for AK 5$|$water we observed non-harmonic wave behavior and wave breaking solely for a few frequencies around the first eigenfrequency $\omega_{11}$ for $d_s \gtrsim 25\, {\rm mm}$. Hence, we can experimentally confirm a general broader applicability of the linear Potential model to interfacial waves than to free-surfaces waves. On the basis of the above findings it could be promising for further research to include damping and interfacial tension into the model. Also modeling the influence of different contact line boundary conditions would be rewarding. 

\section{Conclusion}
We presented a novel interfacial wave experiment allowing to measure orbitally driven wave amplitudes and phase shifts in two- and three-fluid-layer cylindrical containers. It represents a hydrodynamical model for both aluminum reduction cells and liquid metal batteries with the intention to get new insights into the metal pad roll instability. Different inmiscible working liquids are presented and discussed allowing to form various stable two- and three-layer stratifications.\\
Aiming to facilitate measurements of interfacial wave motion also in opaque liquid metals, we introduced a new acoustic measurement procedure. It allows to reconstruct wave amplitudes in the range $0.15\, {\rm mm} \lesssim \eta_0 \lesssim 10\, {\rm mm}$ with a accuracy up to $0.03\, {\rm mm}$. Existing linear theory of orbitally driven one-layer surface waves was extended to two-layer interfacial waves. We justified that the two-layer theory is even of greater practical use than the one-layer theory, since the parameter space of the linear regime is considerably increased for interfacial waves in comparison with free-surface waves. \\
Finally, we presented different measurements of forced interfacial wave motion. Resonance curves and phase shifts between shaker and wave were captured for various interface positions. A large impact of the bottom and top container walls was observed. Amplitudes and eigenfrequencies were considerably decreased for interfaces getting close to the walls. Lastly, we compared the resonance curves with the adapted theory and found a crucial influence of the contact line boundary condition. In cases we realized free-sliding contact lines with a static contact angle close to $90^{\circ}$, the resonance dynamics was well predicted by the theory. In cases, where we observed partially or completely pinning contact lines, the measured resonance frequencies were substantially higher than predicted. This result clearly emphasizes the importance to control the contact line behavior in interfacial wave experiments.

\begin{acknowledgements}
The authors would like to thank Peggy Jähnigen and Kerstin Eckert for the realization of extensive interfacial tension measurements. Fruitful discussions with Kerstin Eckert, Bernd Willers, Norbert Weber and Wietze Herreman on several aspects of acoustic measurement techniques and contact line dynamics are gratefully acknowledged.
\end{acknowledgements}

\bibliographystyle{spbasic}      
\bibliography{literature}   

%
%

\end{document}